\documentstyle[12pt,epsfig]{article}
\newcommand \be{\begin{equation}}
\newcommand \bea{\begin{eqnarray}}
\newcommand \ee{\end{equation}}
\newcommand \eea{\end{eqnarray}}
\newcommand{\lp}{\left(}
\newcommand{\rp}{\right)}

\begin{document}

\title{Comment on A.-L. Barabasi, Nature 435 207-211 (2005)}

\author{Anders Johansen \\
Teglg\aa rdsvej 119, 3050 Humleb\ae k, Denmark}
\date{\today}
\maketitle

The purpose of this communication is twofold. First, it clarifies the origin
of the power law obtained in the computer simulations presented in A.-L. 
Barabasi, Nature 435 207-211 (2005) as well as presenting a statistically more
sound analysis of the experimental email data used in that paper and in 
\cite{paviaemail}. The second purpose is to clarify the origin of the the 
suggestion of power law probability distribution for the response times of 
humans presented with a piece of new information either as a news event or 
through the reception of an email. 

In a letter in Nature published May 2005 \cite{barabasi}, A.-L. Barabasi 
presents results from an email experiment\cite{eck}, which suggest that 
the time lag $\tau$ between the reception of an email and the following 
reply for {\it an individual} follows a power law probability distribution
\be \label{ppl}
P(\tau) \sim \tau^{-1},
\ee
when averaged over time. Only a single example(user) is shown and the claimed
scaling region is extremely limited.

At the conference ``Frontier Science 2003 - A Non-Linear World: The Real 
World'' in Pavia, Italy, Sept. 2003, where A.-L. Barabasi also was present
and more than one year prior to the submission of \cite{barabasi} to Nature, 
I showed with high statistical significance, {\it i.e.}, much higher than that 
of \cite{barabasi}, using {\it the same data set} of \cite{eck}, that when 
{\it averaged over both time and population} the time lag $\tau$ follows a 
power law probability distribution
\be \label{omori}
P(\tau) \sim (\tau + c)^{-1},
\ee
where the reason behind the constant $c$ is simply that the computer time lag
measured is not the true response time\footnote{A quite crude approximation is
made here as the time-lag $c$ is taken to be the same for all individuals. 
This explains the small deviation between the data and the fit for the 
smallest times}. The prime reason for the ``shift'' $c$ is that most 
people do not download and/or read new e-mail messages instantaneously. Not 
only is the statistical significance substantially higher compared with 
\cite{barabasi} because of the population averaging, but also because the 
cumulative distribution of measurements were used thus filtering out much of 
the high-frequency (Integration effectively result in a low-pass filtering). 
The deviation between the data and the logarithmic fit 
for long times can be attributed to to limitations in peoples ability to 
answer a large number of email, i.e., a finite size effect due to a limited 
amount of time and memory, as well as under-sampling for large time lags due 
to the limited population considered. One should note that an alternative 
explanation for the response times of a single individual has been proposed 
using a log-normal distribution \cite{amaral}. My main objection to their 
suggestion, is that I do not know how to interpret the underlying Gaussian 
distributed log(variable).

\begin{figure}[t]
\vspace{-15mm}
\begin{center}
\epsfig{file=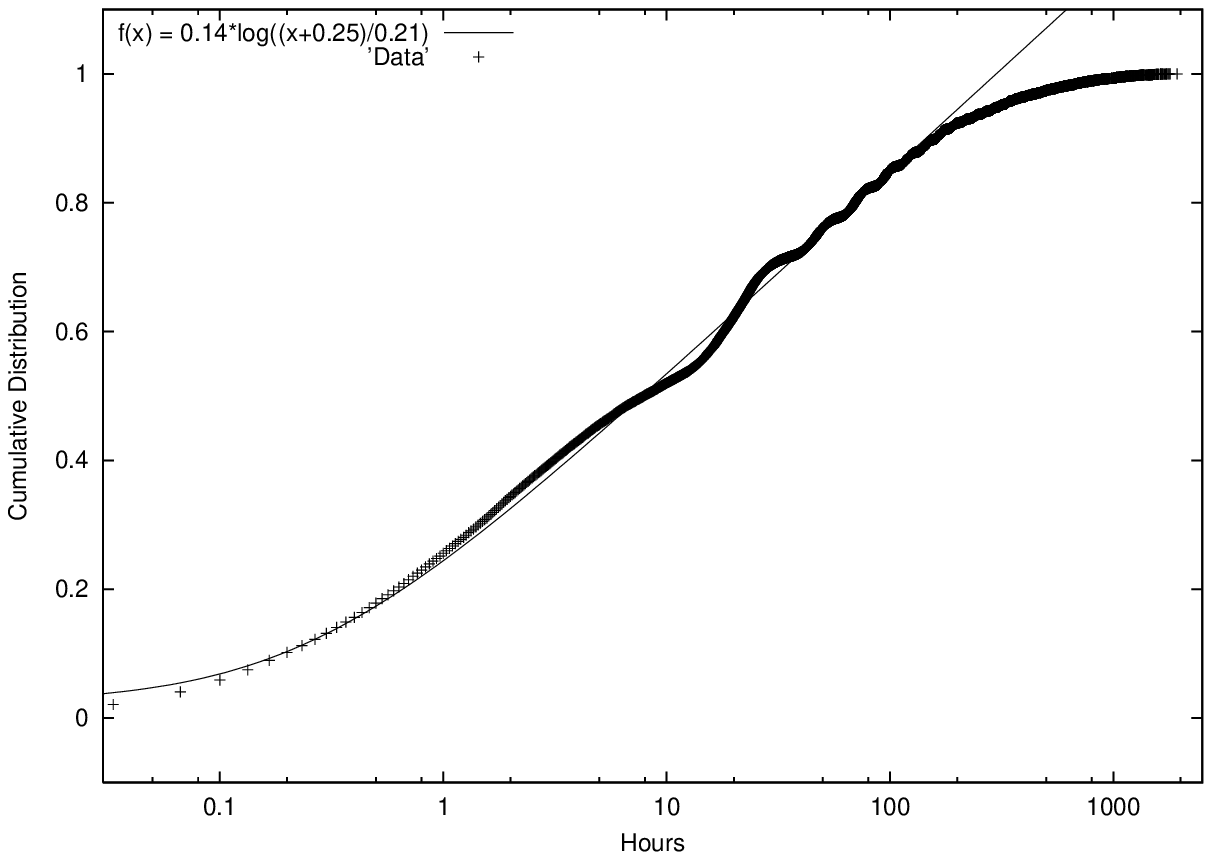,height=8cm,width=13.5cm}
\caption{\protect\label{emailfig1} Cumulative distribution of responses as a
function of time. The fit is $N\lp t\rp = a\ln \lp \lp t + c \rp / b 
\rp$ with  $a\approx 0.14$, $b\approx 0.21$ hours and $c \approx 0.25$ hours. 
Due to the ``wiggles'', the fit has been stabilized by first estimating $c$ 
from the data and then fitting $a$ and $b$ keeping $c$ fixed. The origin of
the ``wiggles'' is simply that people send e-mail messages just before leaving 
their work place. Since people generally share the same working hours (provided
that they live in the same time zone), those messages are not answered before 
the next day. }

\vspace{7mm}

\epsfig{file=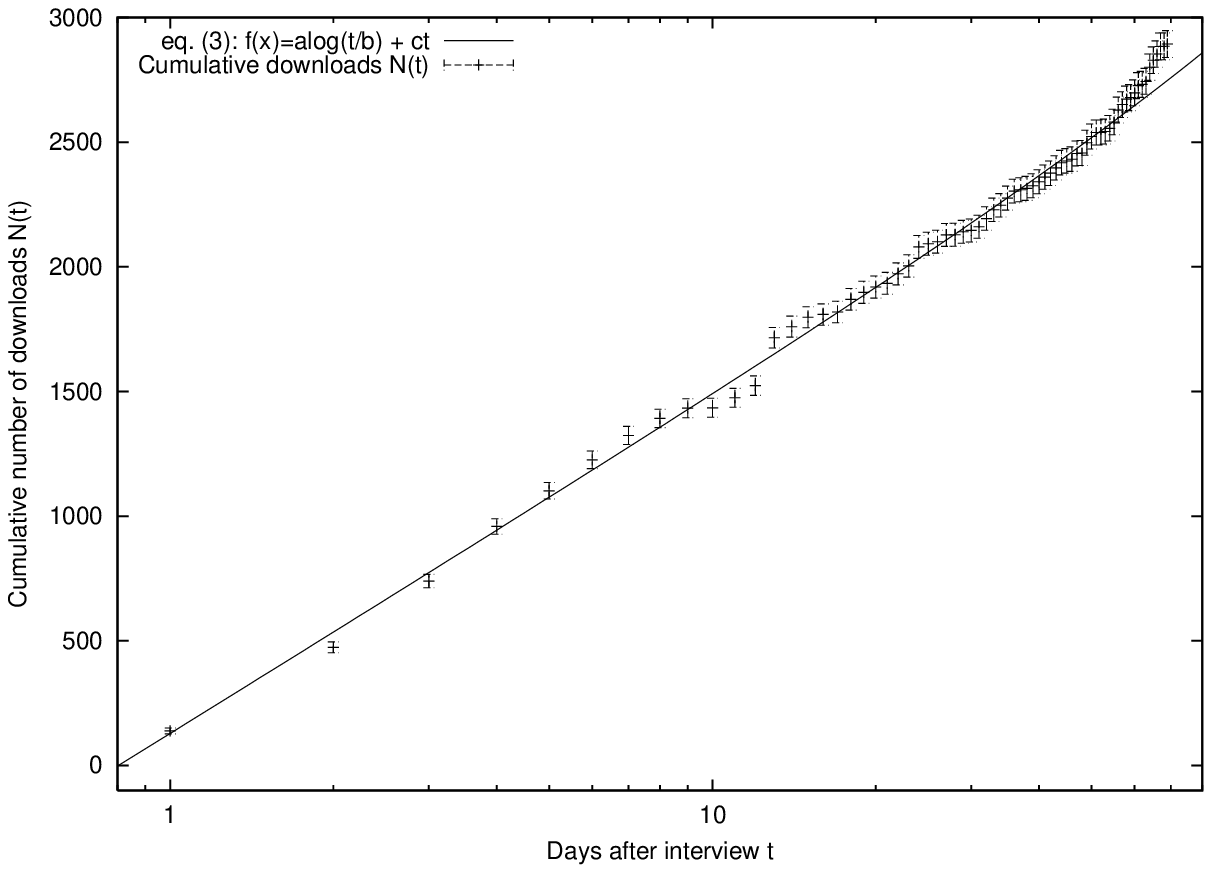,height=8cm,width=13.5cm}
\caption{\protect\label{uncutfig1} Cumulative number of downloads $N\lp t\rp$ 
as a function of time. The fit is $N\lp t\rp = a\ln\lp t/b \rp + kt$ with 
$a \approx 583$, $b \approx 0.80$ days and $k \approx 2.2$ days$^{-1}$. The 
deviation between fit and data after $\sim 60$ days is due to another 
publication of URL on \protect\cite{SSRN}.}
\end{center}
\end{figure}

These findings was later published in the conference proceedings 
\cite{paviaemail} of Feb. 2004, see fig. \ref{emailfig1} for the data 
analysis. Here it was compared with previous results from another Internet 
experiment \cite{www2}, where a portion of the ``internaut population'''s 
response to a forty minute interview with the author on the origin of stock 
market crashes called ``The World (Not) According to GARCH'' was published on 
Friday the 26th of May 2000 on a ``radio website'' called ``Wall Street Uncut''
\cite{uncut}. In this interview, as well as on the website, the URL to the 
author's papers \cite{url} was announced making it clear that work on stock 
market crashes in general and the recent Nasdaq crash in particular could be 
found on the posted URL. The results was that the response to the interview 
and URL publication, measured as the number of downloads of papers from the 
authors homepage as a function of time (days) from the appearance of the 
interview, {\it also} followed a power law probability distribution
\be
P(\tau) \sim \tau^{-1} + k,
\ee
see fig. \ref{uncutfig1} for the data analysis. The constant $k$ is simply a 
``background'' due to downloads from people unaware of the interview as well 
as ``search robots''.  Another experiment of the same type can be found in
\cite{www}, where the sampled time interval is 100 days. Here the exponent 
was found to be $\sim -0.6$ and not $-1$ suggesting that multiple communication
channels might influence the value of the exponent.

Even though the two experiments are not identical, there are a number of 
similarities which establish a correspondence between the two. At any time 
$t$ after the appearance of the interview on \cite{uncut}, the exposed 
population consists of two groups, namely those who have not downloaded a 
paper from \cite{url} and those who have. Similarly with respect to the email 
experiment, at any time $t$ the population considered consists of two groups, 
namely those who have an e-mail to answer and those who have not. In both 
cases, the time lag $\tau = t-t_0$, where $t_0$ is the time of the appearance
of the interview/reception of an email to answer, is the governing variable.
The transition from the first state (no action yet) to the second state
(have downloaded/answered email) demands the crossing of some threshold 
specific to each individual. We thus imagine that the announcement of
the URL/the reception of e-mails plays the role a ``field'' to which the 
exposed population is subjected and study the relaxation process by monitoring 
the number of downloads/the number of replies as a function of time. Hence, we 
may view the process of downloading/replying as a diffusion process in a random
potential, where the act of downloading/replying is similar to that of 
barrier-crossings.

In fact, the queuing model proposed by A.-L. Barabasi in \cite{barabasi} is not
much more than a reformulation of the Trap-model proposed by myself and 
co-author in \cite{www} and subsequent papers \cite{www2,paviaemail} as an 
analog to the experiments. Both models use the ad hoc assumption of a power law
 ``trapping time'' distribution $p(\tau) \propto \tau^\gamma$ and introducing
a ``priority parameter'' \cite{barabasi} does not add much new. With respect 
to his computer simulation, it is well-known that a uniformly random sampling 
of an exponentially distributed random variable will trivially give a power law
with exponent of -1 \cite{reedhughes}, so it is not obvious what his computer 
simulations are suppose to prove.

In the conference proceeding \cite{paviaemail}, I speculate over the origin of 
such power law response times distributions and specifically whether it is a 
consequence of the averaging over a population or whether it over time holds on
the individual level as well. I list a number of purely qualitative arguments
suggesting that this might be the case, but conclude that ``it seems a priori 
a quite formidable task to empirically verify whether these considerations are 
valid or not'' with sufficient statistical significance. In \cite{barabasi},
this problem have not been solved at all.

In conclusion, the {\it only} difference between the experimental results 
suggesting a power law distribution of response times presented 
{\it first} in \cite{paviaemail} and approximately a year later in 
\cite{barabasi}, is that the former employs an ensemble averaging whereas
the latter does not. Considering the quite limited scaling region of figure
2b in \cite{barabasi} as well as the scatter of the points, the author's 
conclusion that eq.(\ref{ppl}) holds for a single individual is not obvious,
but certainly interesting. Compared with this, the scaling region in fig. 
\ref{emailfig1} is over 3 decades. I sincerely hope that A.-L. Barabasi in the 
future will give due credit to reference \cite{paviaemail} and that Nature's 
editors and reviewers in the future will follow standard academic procedures 
for referencing background material. In fact, in 2000, the first experimental 
results on human response times on a news event, specifically that of an 
interview published in one of the leading danish newspapers including the 
author's URL \cite{www}, showing a power law distribution of response times 
was submitted to Nature and rejected on the grounds of ``too many papers on 
the Internet''.
\\

{\bf Acknowledgement} The author would like to thank Aaron Clauset and Luis A. Nunes Amaral.

\newpage

\end{document}